\documentclass[11pt]{article}

\usepackage{amsfonts,amssymb,graphics,epsfig,verbatim,bm,latexsym,amsmath,url,amsbsy,
geometry}

\newtheorem{lemma}{Lemma}
\usepackage[authoryear]{natbib}
\usepackage{setspace}

\bibpunct{(}{)}{;}{a}{,}{,}

\begin{document}
\bibliographystyle{plainnat}

\title{\bf Quantifying the Computational Advantage of Forward Orthogonal Deviations}
\date{August 2018}
\author{Robert F. Phillips\footnote{Address: 2115 G Street NW, Suite 340, Washington DC 20052; Telephone: 202-994-8619;  Email: rphil@gwu.edu}\\Department of Economics\\George Washington University}

\maketitle

\begin{abstract}
Under suitable conditions, one-step generalized method of moments (GMM) based on the first-difference (FD) transformation is numerically equal to one-step GMM based on the forward orthogonal deviations (FOD) transformation.  However, when the number of time periods ($T$) is not small, the FOD transformation requires less computational work. This paper shows that the computational complexity of the FD and FOD transformations increases with the number of individuals ($N$) linearly, but the computational complexity of the FOD transformation increases with $T$ at the rate $T^{4}$ increases, while the computational complexity of the FD transformation increases at the rate $T^{6}$ increases. Simulations illustrate that calculations exploiting the FOD transformation are performed orders of magnitude faster than those using the FD transformation. The results in the paper indicate that, when one-step GMM based on the FD and FOD transformations are the same, Monte Carlo experiments can be conducted much faster if the FOD version of the estimator is used.\\


\end{abstract}

\section{Introduction \label{intro}}

In a seminal paper, \cite{Arellano1991} introduced a now well-known one-step generalized method of moments (GMM) panel data estimator.  The estimator relies on first-differencing the observations --- the first-difference (FD) transformation --- and then applying GMM.   \cite{Bover1995} suggested another transformation could be used. They showed that, under suitable conditions, GMM is invariant to how the data are transformed, and they introduced  the forward orthogonal deviations (FOD) transformation to the panel data literature. 
Moreover, \cite{Bover1995} noted there is a computational advantage to using the FOD transformation when the number of columns in the instrument matrix is large. However, to date, there appears to be no published evidence illustrating how much of a computational advantage is conferred by using the FOD transformation.  

The purpose of this paper is to fill that gap. I show how the computational complexity --- the amount of computational worked required --- of the FD and FOD transformations increase with the length of the time series ($T$) and the number of individuals ($N$). The results reveal that, even when lim$(T/N) = 0$,\footnote{In this case,  one-step GMM, based on all available instruments, has no asymptotic bias \citep{Alvarez2003}.} computational complexity is affected much more by the size of $T$ than by the size of $N$. Furthermore, the FD transformation's computational complexity increases with $T$ at a much faster rate than the rate of increase in the FOD transformation's computational complexity. Consequently, when $T$ is not small, the FOD transformation is computationally faster --- orders of magnitude faster --- than the FD transformation.  A practical implication of this finding is that computationally intensive work, such as Monte Carlo simulations, can be performed much faster by relying on the FOD rather than the FD transformation.

\section{The computational complexity of the FD and FOD transformations \label{FOD}}

In order to compare the computational complexity of the FD and FOD transformations,  a simple case was considered --- the first-order autoregressive (AR(1)) panel data model. The model is
\begin{equation}
y_{it}=\delta y_{i,t-1}+\eta _{i}+v_{it},  \label{ar1_model}
\end{equation}%
with $y_{i0}$ taken to be the first available
observation. If the $v_{it}$s are uncorrelated, then, for instruments, one might use $z_{i1}=y_{i0}$, $\boldsymbol{z}_{i2}^{\prime }=\left(
z_{i1},y_{i1}\right) $, $\boldsymbol{z}_{i3}^{\prime }=\left( \boldsymbol{z}%
_{i2}^{\prime },y_{i2}\right) $, and so on up to $\boldsymbol{z}%
_{i,T-1}^{\prime }=( \boldsymbol{z}_{i,T-2}^{\prime },y_{i,T-2}) $%
. For this choice of instruments, one-step GMM based on the FOD transformation is numerically equivalent to one-step GMM based on the FD transformation (see, e.g., Hayakawa and Nagata, 2016).

However, although numerically the same, the two transformations are not computationally the same.  To see this, consider first one-step GMM estimation of  the AR(1) panel
data model using the FD transformation. The first-difference transformation matrix is
\begin{equation*}
\boldsymbol{D}=\left( 
\begin{array}{ccccc}
-1 & 1 & 0 & \cdots  & 0 \\ 
0 & -1 & 1 & \cdots  & 0 \\ 
\vdots  & \vdots  & \ddots  & \ddots  & \vdots  \\ 
0 & 0 & \cdots  & -1 & 1%
\end{array}%
\right). 
\end{equation*}%
The one-step GMM estimator based on this transformation can be written as follows: Let $\boldsymbol{y}_{i}=\left( y_{i1},\ldots
,y_{iT}\right) ^{\prime }$, $\boldsymbol{y}_{i,-1}=\left( y_{i0},\ldots
,y_{i,T-1}\right) ^{\prime }$, $\boldsymbol{\tilde{y}}_{i}=\boldsymbol{Dy}%
_{i}$, and $\boldsymbol{\tilde{y}}_{i,-1}=\boldsymbol{Dy}_{i,-1}$ ($%
i=1,\ldots ,N$).  Also, let $\boldsymbol{Z}_{i}$ denote a block-diagonal matrix with the
vector $\boldsymbol{z}_{it}^{\prime }$ in its $t$th diagonal block ($t=1,\ldots ,T-1$, $%
i=1,\ldots ,N$). Moreover, let $\boldsymbol{\tilde{s}}=\sum_{i}\boldsymbol{Z}%
_{i}^{\prime } \boldsymbol{\tilde{y}}_{i}$, $\boldsymbol{\tilde{s}}_{-1}=\sum_{i}%
\boldsymbol{Z}_{i}^{\prime } \boldsymbol{\tilde{y}}_{i,-1}$, $\boldsymbol{G}=\boldsymbol{D}\boldsymbol{D}^{\prime}$, and $\boldsymbol{A}_{N}=\sum_{i}\boldsymbol{Z}_{i}^{\prime }\boldsymbol{G}\boldsymbol{Z}_{i}$.
Finally, set $\boldsymbol{\tilde{a}}=\boldsymbol{\tilde{s}}_{-1}^{\prime }\boldsymbol{A}_{N}^{-1}$.
Then the one-step FD GMM estimator is given by%
\begin{equation}
\widehat{\delta }_{D}=\frac{\boldsymbol{\tilde{a}}\boldsymbol{\tilde{s}}}{\boldsymbol{\tilde{a}}\boldsymbol{\tilde{s}}_{-1}}.
\label{GMM1_delta}
\end{equation}%
\citep{Arellano1991}. 

For large $T$, the formula in (\ref{GMM1_delta}) is computationally expensive. A measure of computational cost or work is computational complexity, which is the number of  floating point operations or flops required.\footnote{A flop consists of
an addition, subtraction, multiplication, or division.} The conclusion from counting up the number of flops required to compute $\widehat{\delta}_{D}$ is provided in Lemma \ref{FD_flops}.

\begin{lemma} \label{FD_flops}
Given $\boldsymbol{D}$, $\boldsymbol{y}_{i}$, $\boldsymbol{y}_{i,-1}$ and $\boldsymbol{Z}_{i}$ ($i=1,\ldots,N$), the number of flops required by the FD formula in (\ref{GMM1_delta}) increase with $N$ linearly, for given $T$, and increase with $T$ at the rate $T^{6}$ increases, for given $N$.
\end{lemma}

Appendix A.1 provides the flop counts that verify Lemma \ref{FD_flops}.

Lemma \ref{FD_flops} shows that, even when $T$ is much smaller than $N$, it can be much more important than $N$ in determining the amount of computational work --- and hence time --- it takes to obtain an estimate via the formula in (\ref{GMM1_delta}).  
A substantial contribution to the amount of work required is the computation of $\boldsymbol{A}_{N}$ and then inverting it. Appendix A.1 shows that the number of flops required to calculate $\boldsymbol{A}_{N}$ is on the order of O($NT^{5}$).  Moreover, the work required to invert $\boldsymbol{A}_{N}$ increases even faster with $T$. Standard matrix inversion methods require on the order of another O($T^{6}$) flops to compute $\boldsymbol{A}_{N}^{-1}$. 

On the other hand, the FOD transformation does not require inverting $\boldsymbol{A}_{N}$. This fact makes it more efficient computationally when $T$ is not small. 

To see how much more efficient the FOD transformation is, consider again the AR(1)
model in (\ref{ar1_model}), and set $\boldsymbol{\ddot{y}}_{i}=\boldsymbol{Fy}_{i}$ and $\boldsymbol{\ddot{y}}_{i,-1}=\boldsymbol{Fy}_{i,-1}$, where $\boldsymbol{F}$ is the FOD transformation matrix given by 
\begin{eqnarray*}
\boldsymbol{F} &=&\text{ diag}\left( \left( \frac{T-1}{T}\right)
^{1/2}, \left( \frac{T-2}{T-1}\right)
^{1/2}, \ldots ,\left( \frac{1}{2}\right) ^{1/2}\right)   \notag \\
&& \times \left( 
\begin{array}{ccccccc}
1 & - \frac{1}{T-1} & - \frac{1}{T-1} & \cdots & - \frac{1}{T-1} & - \frac{1}{T-1} & - \frac{1}{T-1} \\ 
0 & 1 & - \frac{1}{T-2} & \cdots & - \frac{1}{T-2} & - \frac{1}{T-2} & - \frac{1}{T-2} \\ 
\vdots & \vdots & \vdots &  & \vdots & \vdots & \vdots \\ 
0 & 0 & 0 &  & 1 & -\frac{1}{2} & -\frac{1}{2} \\ 
0 & 0 & 0 &  &  & 1 & -1%
\end{array}%
\right)  \label{FOD_trans}
\end{eqnarray*}%
(see Arellano and Bover, 1995). 
Also, let $\ddot{y}_{it}$ and $\ddot{y}_{i,t-1}$ denote the $t$th entries in $\boldsymbol{\ddot{y}}_{i}$ and $\boldsymbol{\ddot{y}}_{i,-1}$. Then the FOD version of the one-step GMM estimator is
\begin{equation}
\widehat{\delta }_{F}=\frac{\sum_{t=1}^{T-1}\boldsymbol{\ddot{a}}_{t}%
\boldsymbol{\ddot{s}}_{t}}{\sum_{t=1}^{T-1}\boldsymbol{\ddot{a}}_{t}%
\boldsymbol{\ddot{s}}_{t-1}}.  \label{FOD_ar1}
\end{equation}%
where $\boldsymbol{\ddot{s}}_{t}=\sum_{i}\boldsymbol{z}_{it}\ddot{y}_{it}$, $\boldsymbol{\ddot{s}}_{t-1}=\sum_{i}%
\boldsymbol{z}_{it}\ddot{y}_{i,t-1}$, $\boldsymbol{\ddot{a}}_{t}=\boldsymbol{\ddot{s}}_{t-1}^{ \prime }%
\boldsymbol{S}_{t}^{-1}$, and $\boldsymbol{S}_{t}=\sum_{i}\boldsymbol{z}_{it}%
\boldsymbol{z}_{it}^{\prime }$ (see, e.g., Alvarez and Arellano, 2003).

The formula in (\ref{FOD_ar1}) replaces computing and then inverting one large matrix --- the matrix $\boldsymbol{A}_{N}$ --- with computing and inverting several smaller matrices --- the matrices $\boldsymbol{S}_{t}$ ($t=1,\ldots ,T-1$).  The computational savings of this alternative approach are summarized in  Lemma \ref{FOD_flops}.

\begin{lemma} \label{FOD_flops}
Given $\boldsymbol{F}$, $\boldsymbol{y}_{i}$, $\boldsymbol{y}_{i,-1}$, and $\boldsymbol{Z}_{i}$ ($i=1,\ldots,N$), the number of flops required by the FOD formula in (\ref{FOD_ar1}) increase with $N$ linearly, for given $T$, and increase with $T$ at the rate $T^{4}$ increases, for given $N$.
\end{lemma}

The flop counts are provided in Appendix A.2.

Lemmas \ref{FD_flops} and \ref{FOD_flops} show that the computational complexity of one-step GMM based on both the FOD and FD transformations increases much faster with $T$ than with $N$. But the number of flops increase with $T$ at a much slower rate for the FOD transformation.  This finding indicates that, for large $T$, computing time will be orders of magnitude faster for the FOD transformation than for the FD transformation.  This conjecture is explored in the next section.

\section{An illustration\label{MC}}

In order to illustrate the reductions in computing time from using the FOD
transformation rather than differencing, some experiments were conducted.
For all of the experiments, observations on $y_{it}$ were generated
according to the AR(1) model%
\begin{equation*}
y_{it}=\delta y_{i,t-1}+\eta _{i}+v_{it}\text{, \ \ \ \ \ \ \ \ \ }%
t=-49,\ldots ,T,\ i=1,\ldots ,N,
\end{equation*}%
with $y_{i,-50}=0$. The value for $\delta $ was fixed at $0.5$ for all
experiments. The error components $\eta _{i}$ and $v_{it}$ were generated
independently as standard normal random variables. The processes were started with $t=-50$ so that, for each $i$, the process was essentially stationary by
time $t=0$. As for the sample sizes, $T$ was set to either five, 10, 15, 20,
25, 30, 35, 40, 45, or 50 whereas $N$ was either 100, 200, 300, 400, or 500.
After a sample was generated, start-up observations were discarded so that
estimation was based on the $T+1$ observations $y_{i0},\ldots ,y_{iT}$ for
each $i$ ($i=1,\ldots ,N$). Finally, one-step GMM estimates were
calculated both with the FD formula in (\ref{GMM1_delta}) and the FOD formula in (%
\ref{FOD_ar1}). The estimates obtained from the two formulas were identical
but how long it took to compute them differed.\footnote{All computations were performed using GAUSS. To calculate elapsed times for the FOD and FD algorithms, the GAUSS command hsec was used to calculate the number of hundredths of a second since midnight before and after calculations were executed.}

\begin{figure}[t]
\noindent \text{\textbf{Fig. 1} Computing time for $T = 5$ and $N$ increasing from 100 to 500 ($N =$
100, 200, 300, 400,} \\ \text{and 500) over computing time for $T = 5$ and $N = 100$.} \\
\includegraphics[width=\textwidth, height=110mm]{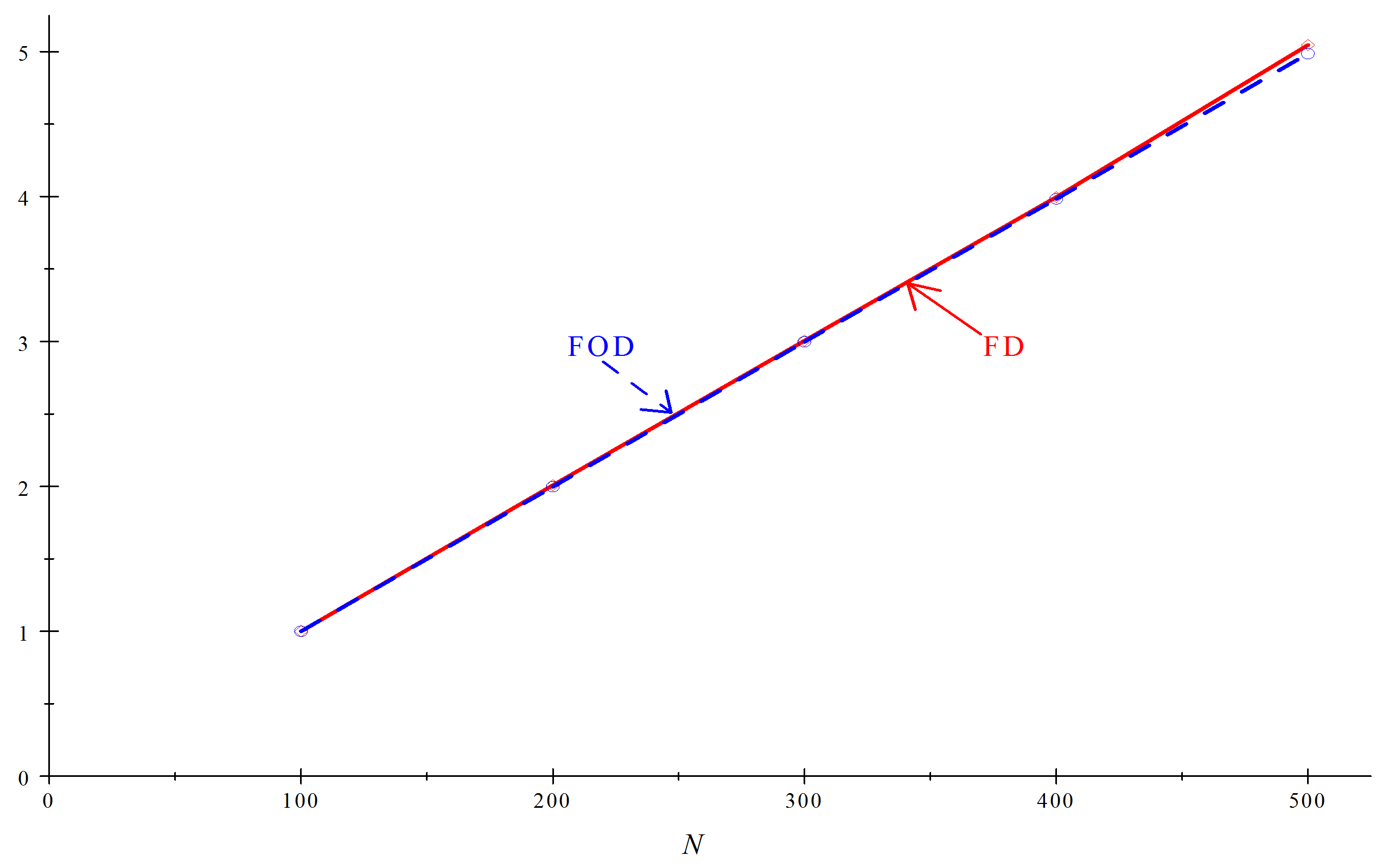}
\end{figure}

Figure 1 plots ratios of computing times as $N$
increases holding $T$ fixed. In Figure 1, I plot the time required to
calculate GMM estimates for 100 independent samples of size $T=5$ and $N=X$ (%
$X=100$, 200, 300, 400, and 500) over the time required to calculate that
many one-step GMM estimates for $T=5$ and $N=100$, first using the FD formula in (\ref{GMM1_delta}) and then using the FOD formula in (\ref{FOD_ar1}). The solid line shows how
computing time increases with $N$, holding $T$ constant, using differencing, while the dashed line
shows how it increases with $N$ using forward orthogonal deviations.%

The message of the figure is clear. Because the number of computations
required to compute a GMM estimate increases linearly in $N$, regardless of whether differencing or the FOD transformation is used, so too does
computing time. For example, in Figure 1, when $N$ is doubled from 100 to
200, computing time approximately doubles, regardless of whether estimates
are computed using the differencing or the FOD transformation. When $N$
triples from 100 to 300, computing time approximately triples, and so on.%

Figure 2 plots ratios of computing times as $T$ changes, with $N$ held fixed
at 100. Specifically, it gives the time required to calculate GMM estimates
for 100 independent samples of size $N=100$ and $T=X$ ($X=5$, 10, 15, 20,
25, 30, 35, 40, 45, and 50) over the time required to compute that many
estimates for $N=100$ and $T=5$. As in Figure 1, the solid curve shows how
computing time increases for the FD transformation, but now as $T$ increases with $N$
held fixed. The dashed curve shows how computing time increases as $T$
increases using the FOD transformation.%

\begin{figure}[t]
\text{\textbf{Fig. 2} Computing time for $N = 100$ and $T$ increasing from 5 to 50 ($T = $ 5, 10, 15, 20, 25, 30,
35,} \\ \text{40, 45, and 50) over computing time for $N = 100$ and $T = 5$.} \\
\includegraphics[width=\textwidth, height=110mm]{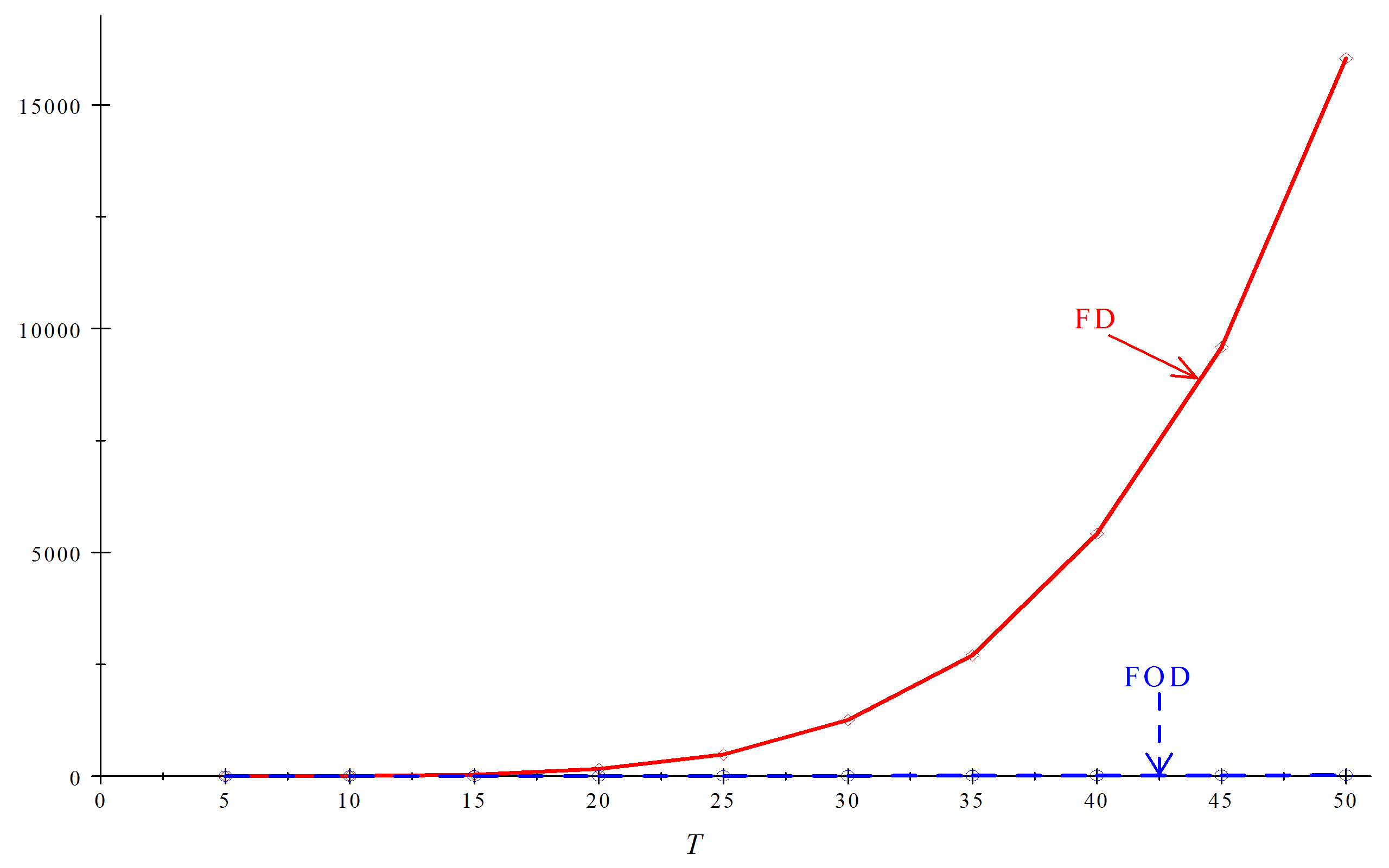}
\end{figure}

Figure 2 shows that computing time does not increase linearly with $T$.
Instead, when the FD transformation is used, computing time increases much faster than linearly. For example, if  the differencing formula in (\ref{GMM1_delta}) is used to calculate one-step GMM
estimates, then, for $N=100$, it takes about 5.5 times longer to calculate
an estimate, on average, when $T=10$ than when $T=5$. Thus, doubling $T$
increases computing time by 5.5 times. When $T$ increases from 5 to 15, a
three-fold increase in $T$, it takes about 36 times longer to calculate an
estimate. Finally, if we increase $T$ ten-fold, from 5 to 50, it takes
approximately $16,038$ times longer to compute GMM estimates if differencing
is used (see Figure 2).

When we use the FOD formula in (\ref{FOD_ar1}) to calculate GMM
estimates, the relative increase in computing time is much less dramatic,
but it is still not linear in $T$. In Figure 2, I also plotted the ratios
of computing times required to calculate GMM estimates using the FOD formula in (\ref%
{FOD_ar1}). For the FOD transformation, how relative computing time
increases with $T$ is indicated by the dashed line in the figure. The line
hugs the horizontal axis in Figure 2 because the increase in computing time
with $T$ is of a much smaller order of magnitude when the FOD transformation
is used than when the FD transformation is used. Specifically, for $N=100$, the computing time given $T=10$ is 2.2
times what it is when $T=5$. Thus, doubling $T$ leads to a bit more than double the computing time. However, a ten-fold increase in computing time --- when $T$ is increased from 5 to 50 --- leads to computing time taking about 24 times longer to compute
an estimate when the FOD transformation is used. 

Although the computational complexity of the FOD transformation is
not linear in $T$, it increases at a much slower rate in $T$ than is the
case when differencing is used. Consequently, as $T$ increases, using the
FOD formula for calculating GMM estimates leads to
significant reductions in computing time relative to using the differencing
formula for computing GMM estimates. Table 1 shows how large these
reductions can be.

Table 1 reports  time ratios. The ratios in Table 1 are the time required to compute estimates using the
FD transformation over the time required to compute those same
estimates using the FOD transformation for different values of $T$ and $N$.
For small $T$, the computations based on first differences are faster. For
example, for $T=5$, it takes about half the time to calculate an
estimate using the FD formula in (\ref{GMM1_delta}) rather than the FOD formula in (%
\ref{FOD_ar1}). However, because computational complexity increases in $T$
at a slower rate when the FOD transformation is used rather than
differencing, the FOD method is faster --- indeed,
much faster --- than differencing for larger values of $T$. For $T$ as small as 10, the FOD transformation is faster, and, for 
$T=50$, computations based on the FOD transformation are two orders of magnitude faster --- specifically, over 300 times faster --- than computations exploiting the FD transformation.

\vspace{10mm}

Table 1: Time to compute FD GMM estimates over time to compute FOD GMM estimates.

\begin{tabular}[h]{lccccccccccc}

\hline
&  &  &  &  &  &  &  &  &  &  &  \\ 
&  & \multicolumn{10}{c}{$T$} \\ 
&  & $5$ & $10$ & $15$ & $20$ & $25$ & $30$ & $35$ & $40$ & $45$ & $50$ \\ 
\hline
\multicolumn{1}{c}{} &  &  &  & \multicolumn{1}{r}{} &  &  &  &  &  &  &  \\ 
$N=100$ &  & 0.47 & 1.18 & 4.82 & 15.24 & 33.30 & 65.05 & 106.93 & 167.56 & 
235.44 & 316.90 \\ 
$N=200$ &  & 0.47 & 1.20 & 4.86 & 15.62 & 33.62 & 61.60 & 104.07 & 162.72 & 
236.71 & 319.03 \\ 
$N=300$ &  & 0.47 & 1.18 & 4.92 & 15.89 & 33.64 & 61.70 & 104.19 & 163.38 & 
237.45 & 320.49 \\ 
$N=400$ &  & 0.47 & 1.19 & 5.19 & 15.97 & 33.89 & 61.55 & 104.39 & 163.49 & 
240.04 & 321.96 \\ 
$N=500$ &  & 0.48 & 1.20 & 5.24 & 15.90 & 33.66 & 61.98 & 104.63 & 163.48 & 
239.43 & 321.65 \\ 
&  &  &  &  &  &  &  &  &  &  &  \\ \hline 

\end{tabular}%

Note: Each estimate is based on 100 samples.

\section{Summary and concluding remarks}

This paper showed that the computational complexity of one-step GMM estimation based on the FD transformation increases with $N$ linearly but it increases with $T$ dramatically --- at the rate $T^{6}$ increases. On the other hand, when the FOD transformation is used, computational complexity still increases with $N$ linearly, but it increases with $T$ at the rate $T^{4}$ increases. Simulation evidence provided in Section \ref{MC} showed that the reductions in computing time from use of the FOD instead of the FD transformation are dramatic when $T$ is not small.

The fact that estimates can be computed so much faster with the FOD transformation implies that Monte Carlo simulations, and other computationally intensive procedures, can be conducted in a fraction of the time if the FOD transformation is used rather than the FD transformation.  Consequently, Monte Carlo studies using large values of $T$ and complicated models that may be prohibitively costly for GMM estimation based on the FD transformation may be feasible for GMM based on the FOD transformation.

\section*{Appendix A: Floating point operations for one-step GMM}

A floating point operation (flop) is an addition, subtraction, multiplication, or division. This appendix shows how the number of flops 
required to calculate $\widehat{\delta }$ via the formulas in (\ref%
{GMM1_delta}) and (\ref{FOD_ar1}) depends on $N$ and $T$. 

To find the number of
flops required to calculate $\widehat{\delta }$, the following facts will be
used repeatedly throughout this appendix. Let $\boldsymbol{B}$, $\boldsymbol{%
E}$, and $\boldsymbol{H}$ be $q\times r$, $q\times r$, and $r\times s$
matrices, and let $d$\ be a scalar. Then $d\boldsymbol{B}$, $\boldsymbol{B}%
\pm \boldsymbol{E}$, and $\boldsymbol{BH}$ consist of $qr$, $qr$, and $%
qs \left( 2r-1\right) $ flops, respectively (see Hunger, 2007).%

\subsection*{Appendix A.1: Floating point operations using differencing}

 To calculate $\boldsymbol{\tilde{y}}_{i} =\boldsymbol{D}\boldsymbol{y}_{i}$ ($i=1,\ldots,N$), a total of  $N\left(T-1\right)\left(2T-1\right)$ flops are needed. After the $\boldsymbol{\tilde{y}}_{i}$s are calculated, the number of flops required to compute $\boldsymbol{\tilde{s}}%
=\sum_{i}\boldsymbol{Z}_{i}^{\prime } \boldsymbol{\tilde{y}}_{i}=\left( 
\boldsymbol{Z}_{1}^{\prime },\ldots \boldsymbol{Z}_{N}^{\prime }\right)
\left(  \boldsymbol{\tilde{y}}_{1}^{\prime },\ldots , \boldsymbol{\tilde{y}}%
_{N}^{\prime }\right) ^{\prime }$ is $m\left[ 2N\left( T-1\right) -1\right] $,
 where $m=T (T -1) /2$ is the number of moment restrictions. Therefore, the total number of flops required to calculate $\boldsymbol{\tilde{s}}$ is $N\left(T-1\right)\left(2T-1\right)+m\left[ 2N\left( T-1\right) -1\right]$. Given $m$ increases with $T$ at a quadratic rate, the number of flops required to compute $\boldsymbol{\tilde{s}}$ is therefore of order O($NT^{3}$). The same number of flops is needed to compute $\boldsymbol{\tilde{s}}_{-1}$. Hence, the number of flops needed to compute $\boldsymbol{\tilde{s}}$ and $\boldsymbol{\tilde{s}}_{-1}$ increase with $N$ linearly, for given $T$, and with $T$ at a cubic
rate, for given $N$.%

To compute $\boldsymbol{A} _{N}$ we must compute $\boldsymbol{G}=\boldsymbol{D}\boldsymbol{D}^{\prime}$, which requires $\left(T-1\right)^{2}\left(2T-1\right)$ flops; the products $\boldsymbol{GZ}_{i}$ $%
\left( i=1,\ldots ,N\right) $, which requires another $Nm\left( T-1\right) \left(
2T-3\right) $ flops; and the products $\boldsymbol{Z}_{i}^{\prime }\left( 
\boldsymbol{GZ}_{i}\right) $ $\left( i=1,\ldots ,N\right) $, which require $Nm^{2}\left( 2T-3\right) $ flops. Finally, we
execute $N-1$ summations of the $m\times m$ matrices $\boldsymbol{Z}%
_{i}^{\prime }\boldsymbol{GZ}_{i}$ $\left( i=1,\ldots ,N\right) $ for
another $\left(N-1\right)m^{2}$ flops. From this accounting, we see that $\left(T-1\right)^{2}\left(2T-1\right)+Nm\left( T-1\right) \left(
2T-3\right) + Nm^{2}\left( 2T-3\right)+\left(N-1\right)m^{2}$
 flops are required to compute $\boldsymbol{A} _{N}$. Given $m$ is quadratic in $T$, the number of flops required to compute $\boldsymbol{A} _{N}$ is of order O($NT^{5}$). Hence, the number of flops increase with $N$ linearly, for given $T$, but they
increase with $T$ at the rate $T^{5}$, for given $N$.%

The number flops required to compute $\boldsymbol{A} _{N}^{-1}$ increases with $%
T$ at the rate $T^{6}$. To see this, note that standard methods for inverting a $q\times q$ matrix require on the order of $q^{3}$
operations (see Hunger, 2007; Strang, 2003, pp. 452--455). The matrix $\boldsymbol{A}
_{N}$ is $m\times m$, and $m$ increases with $T$ at the rate $T^{2}$ if all
available moment restrictions are exploited. Hence, the number of flops required to invert $\boldsymbol{A} _{N}$ is of order O($T^{6}$).

No additional calculations increase with $T$ and $N$ as quickly as computing $\boldsymbol{A}_{N}$ and its inversion. For example, after $\boldsymbol{A}
_{N}^{-1}$ is calculated, $m\left( 2m-1\right) $ flops are required to calculate %
$\boldsymbol{\tilde{a}}=\boldsymbol{\tilde{s}}_{-1}^{\prime }\boldsymbol{A} _{N}^{-1}$, while computing $%
\boldsymbol{\tilde{a}}\boldsymbol{\tilde{s}}_{-1}$, and $\boldsymbol{\tilde{a}}\boldsymbol{\tilde{s}}$ both require $2m-1$ flops.%

\subsection*{Appendix A.2: Floating point operations using FOD}

Calculation of $\boldsymbol{\ddot{y}}_{i} =\boldsymbol{F}\boldsymbol{y}_{i}$ ($i=1,\ldots,N$) requires $N\left(T-1\right)\left(2T-1\right)$ flops. An additional $t\left(
2N-1\right) $ flops are needed to calculate $\boldsymbol{\ddot{s}}_{t}=\left( \boldsymbol{z}_{1t},\ldots ,
\boldsymbol{z}_{Nt}\right) \left( \ddot{y}_{1t},\ldots ,\ddot{y}_{Nt}\right) ^{\prime }$. Therefore, calculation of all of the $\boldsymbol{\ddot{s}}%
_{t}$s ($t=1,\ldots ,T-1$) requires $\ddot{f}_{1}=N\left(T-1\right)\left(2T-1\right)+\left(
2N-1\right) \sum_{t=1}^{T-1}t=N\left(T-1\right)\left(2T-1\right)+\left( 2N-1\right) T\left( T-1\right) /2$ flops, which is of order O($NT^{2}$). Calculation of $\boldsymbol{\ddot{s}}_{t-1}$ ($t=1,\ldots ,T-1$)
requires another $\ddot{f}_{2}=\ddot{f}_{1}$ flops.

\sloppy
On the other hand, computing $%
\boldsymbol{S}_{t}=\left( \boldsymbol{z}_{1t},\ldots ,
\boldsymbol{z}_{Nt}\right)\left( \boldsymbol{z}_{1t},\ldots ,
\boldsymbol{z}_{Nt}\right)^{ \prime}$ requires $t^{2}\left( 2N-1\right) $ flops. Therefore,
calculation of $\boldsymbol{S}_{t}$ ($t=1,\ldots ,T-1$) requires $%
\ddot{f}_{3}=\left( 2N-1\right) \sum_{t=1}^{T-1}t^{2}=\left( 2N-1\right)
T\left( 2T-1\right) \left( T-1\right) /6$ flops, which is of order  O($NT^{3}$). 

\fussy
The matrix $\boldsymbol{S}%
_{t} $ is a $t\times t$ matrix, which requires on the order of O($t^{3}$) flops to invert. Given there are $T-1$ $\boldsymbol{S}_{t}$ matrices
that must be inverted, the number of operations required to invert all of them is on the order of $\ddot{f}_{4}=%
\sum_{t=1}^{T-1}t^{3}=T^{2}\left( T-1\right) ^{2}/4$ flops. In other words, the number of flops required to invert all of the $\boldsymbol{S}_{t}$ matrices is of order O($T^{4}$). 

After $\boldsymbol{S}_{t}^{-1}$ ($t=1,\ldots
,T-1 $) are computed, computing $\boldsymbol{\ddot{a}}_{t}=\boldsymbol{\ddot{s}}%
_{t-1}^{ \prime }\boldsymbol{S}_{t}^{-1}$ ($t=1,\ldots ,T-1$) requires
another $\ddot{f}_{5}=\sum_{t=1}^{T-1}t\left( 2t-1\right) =T\left(
T-1\right) \left(4T-5\right) /6$ flops, which is of order O($T^{3}$). Next, calculation of $\boldsymbol{\ddot{a}}%
_{t}\boldsymbol{\ddot{s}}_{t}$ ($t=1,\ldots ,T-1$) requires $%
\ddot{f}_{6}=\sum_{t=1}^{T-1}\left( 2t-1\right) =T\left( T-2\right) +1$
flops, and then summing the computed $\boldsymbol{\ddot{a}}_{t}\boldsymbol{\ddot{s}%
}_{t}$s --- i.e., $\sum_{t=1}^{T-1}\boldsymbol{\ddot{a}}_{t}%
\boldsymbol{\ddot{s}}_{t}$ --- is another $\ddot{f}_{7}=T-2$ flops. 

Hence, calculation of $\sum_{t=1}^{T-1} \boldsymbol{\ddot{a}}_{t}\boldsymbol{\ddot{s}}_{t}$ requires $%
\sum_{j=1}^{7}\ddot{f}_{j}$ flops.  This work increases with $N$ at the rate $N$
increases, for given $T$, and increases with $T$ at the rate $T^{4}$ increases, for given $N$. 

Of course, to compute $\widehat{\delta }_{F}$ we must also compute $\sum_{t=1}^{T-1}%
\boldsymbol{\ddot{a}}_{t}\boldsymbol{\ddot{s}}_{t-1}$, but the $%
\boldsymbol{\ddot{a}}_{t}$s and $\boldsymbol{\ddot{s}}_{t-1}$s have already been calculated.  Therefore, the
remaining calculations required to compute $\widehat{\delta }_{F}$ are but a small part of the total
number of flops required.

\bibliography{references}

\end{document}